\documentclass[DIV12]{scrartcl}
\usepackage{amsmath}
\usepackage{graphicx}
\usepackage{authblk}
\usepackage[]{SIunits}
\usepackage{cite}
\setkomafont{disposition}{\normalcolor\rmfamily}
\setkomafont{caption}{\small\mdseries\selectfont}

\setcapindent{0mm}%
\newcommand{\Fig}[1]{Fig.~#1}
\newcommand{\Eq}[1]{Eq.~(#1)}
\newcommand{\ndb}{NDB}
\newcommand{\e}[1]{e^{#1}}
\newcommand{\Rect}{\Pi}
\newcommand{\Hexa}{X}
\newcommand{\Ratch}{\Lambda}
\newcommand{\vect}[1]{\textbf{#1}}
\begin{document}
\title{\huge Multiplexing complex two-dimensional photonic superlattices}
\author{\large Martin Boguslawski, Andreas Kelberer, Patrick Rose, and Cornelia Denz}
\affil{\normalsize Institut f\"{u}r Angewandte Physik and Center for Nonlinear Science (CeNoS), \\Westf\"{a}lische Wilhelms-Universit\"{a}t M\"{u}nster, 48149 M\"{u}nster, Germany}
\date{}
\maketitle
\begin{abstract}
\small We introduce a universal method to optically induce multiperiodic photonic complex superstructures bearing two-dimensional (2D) refractive index modulations over several centimeters of elongation. These superstructures result from the accomplished superposition of 2D fundamental periodic structures. To find the specific sets of fundamentals, we combine particular spatial frequencies of the respective Fourier series expansions, which enables us to use nondiffracting beams in the experiment showing periodic 2D intensity modulation in order to successively develop the desired multiperiodic structures. We present the generation of 2D photonic staircase, hexagonal wire mesh and ratchet structures, whose succeeded generation is confirmed by phase resolving methods using digital-holographic techniques to detect the induced refractive index pattern.
\end{abstract}
\section{Introduction}
Inspired by numerous prototypes in nature, photonic structures have become a captivating research subject of great demand and luminous future perspective. Due to excellent theoretical as well as experimental results in the last decades, this field already educed many basic modules for the future technology as to telecommunication and information industry, all-optical computing as well as displaying devices. 

The groundbreaking fundamental insight is the existence of structure-specific band structures \cite{BandStructures}, for instance providing complete band gaps that allow for particular Bloch modes to explicitly be reflected. In analogy to charge carrier propagation processes in condensed matter, many new effects of light propagation in refractive-index modulated photonic structures besides Bragg diffraction \cite{BraggReflection} were discovered, e.g. Bloch oscillation, Zener tunneling, and Anderson localization in the linear \cite{BlochOscillation, BlochZener, AndersonLocalPeriod}, as well as spatial discrete bright, dark, and vortex solitons in the nonlinear regime \cite{discSoliton1, discSoliton2a, discSoliton2b, vortSoliton1, vortSoliton2, multVortSoliton1, multVortSoliton2}. 

Up to now, relatively simple systems were investigated, however the scientific interest aspires at the investigation of more complex systems such as quasiperiodic \cite{quasiperiodPhotonStructures1, quasiperiodPhotonStructures2}, curvilinear including Bessel and Mathieu lattices \cite{BesselSolitons1a, BesselSolitons1b, BesselSolitons1c,BesselSolitons2,MathieuSolitons} where already solitary structures have been found, as well as randomized photonic structures \cite{AndersonLocalization}. One of the most interesting field of highly promising systems regarding spectacular propagation effects are multiperiodic photonic structures, as especially in these systems intriguing analogies of quantum mechanics such as Klein Tunneling or Zitterbewegung could be predicted theoretically as well as demonstrated experimentally \cite{KleinTunneling, Zitterbewegung1, Zitterbewegung2}, not disregarding distinct solitary solutions \cite{superlattSoliton}.

In addition to direct-laserwritten \cite{directLWrite} or holographic-lithographically \cite{lithography} generated photonic structures, photorefractive materials represent a highly appropriate system to carry out linear as well as nonlinear light propagation experiments \cite{prMedium1, prMedium2, discSoliton2a, discSoliton2b}. The main advantage here is the reconfigurable and dynamically manipulable induction of complex two-dimensional refractive index modulations and thus, the optical induction of periodic, quasiperiodic and curvilinear two-dimensional \cite{optInductionNdBs} as well as three-dimensional photonic structures could be presented \cite{optInd3DLatt}. Additionally, probing of the generated structures is feasible in one and the same setup and no further treatment of the sample is needed. 

The implementation of all the mentioned photonic structures were limited to one individual structure-inducing wave field, that is generated by interferometry, diffraction on artificial holograms, or via wave field moulding per spatial light modulators. As it is advantageous to implement wide-elongated photonic structures offering a longer interplay length which offers an analogue of long distance propatation to temporal behavior, one is constricted to use writing wave fields that are translation invariant in the direction of propagation, represented by the class of nondiffracting beams \cite{ndBs1, ndBs2, ndBs3}. Though, designing wave fields of multiple transverse periodic lengths by coherently summing up various nondiffracting wave fields each showing different structural sizes, the translation invariance of the intensity distribution of each wave is no longer conserved due to occuring longitudinal interference modulations. Anyhow, to implement multiperiodic structures by optical induction using nondiffracting beams, one has to care for an incoherent superposition of several writing beams of various structural sizes, for instance by holographic multiplexing techniques that are well known from the field of holographic data storage \cite{MuXDataStorage, incrMuX}. The technique of incremental multiplexing to modulate the refractive index of a photosensitive medium was presented earlier \cite{holoMuX, ratch1D} where the former publication already presented the multiplexing method using 2D square lattices with different lattice periods as a proof of principle, and the latter demonstrates the generation of a one-dimensional (1D) photonic ratchet structure.

In this contribution, we propose an approach for the optical induction of two-dimensional multiperiodic photonic structures by incremental multiplexing techniques. The outline of the paper is as follows. After this first introductory section, Section \ref{sec:mpStruct} discusses theoretically the specification to create three representative multiperiodic structures. We respond to the proper incoherent combination of fundamental periodic lattices of different structural size to expand the desired superlattices. Recapitulatory, the mathematical considerations of each structure are summarized graphically. The experimental implementation of the developed series expansion is subject of Section \ref{sec:expImplmnt}, where the experimental setup is presented and details of the optical incremental multiplexing method are specified. In Section \ref{sec:analysis} the analysis of the generated photonic superstructures and the results are detailed.

\section{Multiperiodic structures and set of their fundamentals} \label{sec:mpStruct}
This Section addresses the derivations of the 2D superlattices in terms of Fourier series expansions, which can be later applied for the experimental part of optical induction. Therefore, we design the multiperiodic refractive index distribution by a stepwise and incremental induction of single index expansion terms using periodic nondiffracting beams (\ndb s) with varying periodicity as the fundamentals. The usage of \ndb{}s is essential, and thus we exclusively concentrate on this class of light waves as lattice-inducing wave fields and on their transverse intensity distributions in particular. Although there is a wide spectrum of different intensity modulations among the four families of nondiffracting beams, we solely operate during the induction process with periodic discrete \ndb s \cite{vortexBeams, discNondiffBeams} to preserve the periodic character of our photonic structures of inquiry. 

An intuitive approach to generate 2D structures with multiperiodic character is to use an optical Fourier series expansions via 1D basis lattices in different orientations and varying periodic length. In this context, a 2D multiperiodic structure $M$ is designed by combining 1D lattice structures of equal modulation frequency resulting in a set of 2D fundaments of varying periodicity, which accounts for a series expansion of the desired structure. This procedure will for all of our examples end up in different sets of ten 2D basis lattice beams, where the series expansion is symbolically written as
\begin{eqnarray}\label{eq:expSeriesGeneral}
M(\vect r) = \sum_{m = 1}^{10} a_m I(\vect k_m^j \vect r).
\end{eqnarray}
Here, $a_m$ is the weighting coefficient of the $m$th order term and $I$ describes a 2D basis intensity.
Further, the vector $\vect r = (x, y)^T$ characterizes a point in 2D space and $\vect k_m^j$ with $j = 1, 2$ (or depending on the rotational symmetry of the structure $j = 1,2,3$, respectively) represents a modulation vector in the corresponding 2D Fourier space. In general, all following field distributions and their intensities are specified as 2D functions exclusively in a transverse plane. The thereto orthogonal direction $z$ can be identified with the direction of propagation. The field evolution in this direction as well as the temporal development are consequently neglected in the following considerations, as these behaviors can be examined by a multiplication of the field distribution of interest with the term $\exp(i(k_{\|} z - \omega t ))$, where $k_{\|} = (k_\text{3D}^2 - k^2)^{1/2}$ is the coordinate of the 3D wave vector $k_\text{3D}$ in the direction of propagation and $k$ describes the modulation in transverse direction.

In the following, we introduce three representative multiperiodic structures with inherently different properties. Thereby, we concentrate on considerations of intensity distributions since with our technique exclusively the intensity is relevant for refractive index changes caused by the photorefractive effect. To achieve particular intensity distributions, we design corresponding light wave distributions showing nondiffracting properties as they can be very flexibly implemented experimentally \cite{discNondiffBeams, optInductionNdBs}. Since the presented structures are only representative examples, various combinations of rotational symmetry (even quasiperiodic) and expansion rules for arbitrary periodic functions are implementable and, moreover, combinations of different expansion rules for unequal transverse directions are possible. In principle, all 2D intensity distributions are implementable that can be resembled in a series expansion of harmonic functions. Nonetheless, an additional constant offset contribution can occur that might limit the resulting contrast of the structure in some cases.

\subsection{Staircase superlattice} \label{ssec:2DRectStruct}
The first structure we consider is a 2D staircase refractive index superlattice. This unique structure offers steep slopes and flat plateaus, involving high spatial frequencies of nonnegligible significance. For that reason, a successful implementation of such a challenging structure proofs the high functionality and flexibility of the presented technique.

\begin{figure}[t]
	\centering	
	\includegraphics[width=1.00\textwidth]{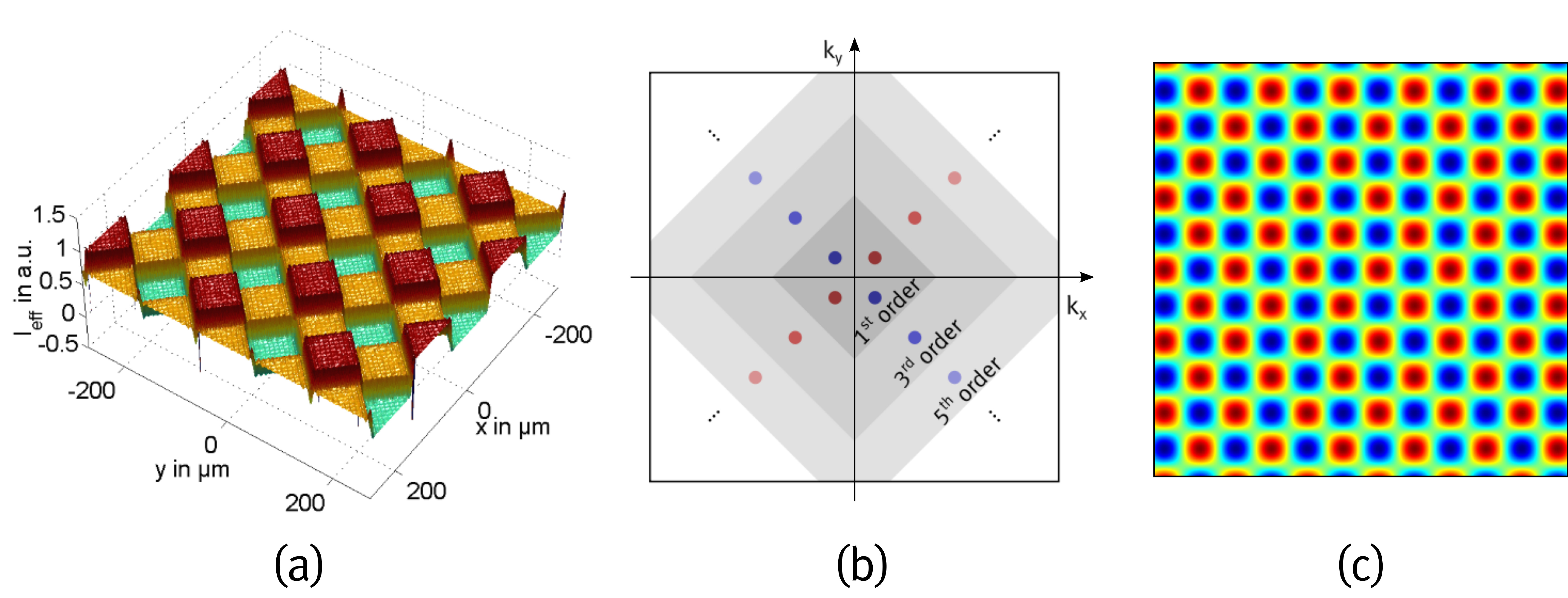}

	\caption{Schematics describing a multiperiodic 2D staircase lattice; (a): surface plot of the effective intensity calculated by series expansion up to an order of ten, (b): presentation of the relevant spatial frequencies, (c) periodic square lattice as the basis structure of each expansion term.}
	\label{fig:Fourier_staircase}
\end{figure}

Figure \ref{fig:Fourier_staircase}(a) represents graphically the desired 2D Fourier series expansion. In the illustrations of the relevant Fourier components sketched in \Fig{\ref{fig:Fourier_staircase}}(b), equally colored spots of a certain brightness represent the sine or cosine functions of one of the orientations. The first expansion orders are represented in gray boxes. Especially in these schematics, the multiperiodic character becomes obvious due to the higher order harmonics. Combinations of all contributions of one order $m$ lead to the derived series expansions with 2D basis sets consisting of nondiffracting beam intensities. 

To derive the correct intensity distribution of each set of 2D fundamentals, we first analyze the expansion rule for a 2D rectangular function $\Rect_{2D}$ with respect to 1D fundamentals. To keep the intensity non-negative, we adjoin an offset term:
\begin{eqnarray}\label{eq:rectangular1D}
\Rect_{2D}(\vect r) = \sum_{m = 1}^{10} \frac{1}{2m-1} \left[2 + \sin\left(\vect k_{m}^1\vect r\right) + \sin\left(\vect k_m^2\vect r\right)\right],
\end{eqnarray}
where $m$ determines the order of the expansion term. Thus, for every expansion term $m$ two $sin$-modulations with orthogonal modulation vectors $\vect k_m^1$ and $\vect k_m^2$ contribute to the total 2D distribution $\Rect_{2D}$. The length of the modulation vectors are given by the arbitrarily chosen lattice period of the first order $g_1$ and the order $m$, hence:
\begin{eqnarray}\label{eq:rect_k_vectors}
\vect{k}_{m}^j = (2m - 1)\;\frac{2\pi}{g_1} \; \vect{e}_j = k_m\ \vect{e}_j \qquad \text{with}\  j = 1,2,\quad \vect{e}_1 = (1, 1),\  \vect{e}_2 = (1, -1).
\end{eqnarray}
In the next step we will combine the 1D intensity contributions of a particular value of $m$. These terms reveal modulation vectors $\vect k_m^j$ which are transverse components of wave vectors with mutual equal projection length parallel to the direction of propagation and thus restrain any modulation of intensity in this direction which is characteristic for \ndb s. The sum of the two sinusoidal intensity contributions for a fixed $m$ in \Eq{\ref{eq:rectangular1D}} can be rewritten as
\begin{eqnarray}\label{eq:sinTerms}
\begin{split}
\left[2 + \sin\left(2\phi_1\right) + \sin\left(2\phi_2\right)\right]&\\ 
=& \frac{1}{2}\left[4 + \e{i(2\phi_1 - \pi/2)} + \e{-i(2\phi_1 - \pi/2)} + \e{i(2\phi_2 - \pi/2)} + \e{-i(2\phi_2 - \pi/2)}\right]\\
=& \frac{1}{2}\left|\e{i(\phi_1 - \pi/4)} + \e{-i(\phi_1 - \pi/4)} + i\e{i(\phi_2  - \pi/4)} + i\e{-i(\phi_2  - \pi/4)}\right|^2,
\end{split}
\end{eqnarray}
where we introduced the phase function $\phi_j$ for a particular $m$ with $j = 1,2$ using $2\phi_{j, m} = \vect{k}_m^j \vect r$ and dropping the index $m$. Now it becomes obvious that the wave field whose absolute square value appears in \Eq{\ref{eq:sinTerms}} equals the intensity of a field distribution consisting of four interfering plane waves. This field belongs to the four-fold discrete \ndb s, whose transverse intensity distribution is presented in \Fig{\ref{fig:Fourier_staircase}}(c) and which is referred as $\Psi_{4,1}(\vect r)$ in Ref. \cite{discNondiffBeams}. In the case here, every plane wave carries an additional phase shift of $\pm \pi/4$, where plane waves of opposite spatial frequency have opposite sign, resulting in a real space translation of the structure in transverse direction (cf. Ref. \cite{discNondiffBeams}).

Hence, we can transcribe \Eq{\ref{eq:rectangular1D}} to
\begin{eqnarray}\label{eq:rectangular2D}
\Rect_{2D}(\vect r) = \sum_{m = 1}^{10} \frac{1}{2(2m-1)} {\left|\Psi_{4,1}(\vect r, k_m)\right|^2}
\end{eqnarray}
and receive a series expansion of a multiperiodic 2D staircase structure incorporating transverse intensities of a nondiffracting wave field as the fundamental structures.

\subsection{Hexagonal wire mesh superlattice}
In this Subsection we derive an expression for the set of basis lattices creating a hexagonal multiperiodic structure. Due to their hexagonal property such systems are of particular importance and coveted subject of many investigations for instance in graphene physics \cite{graphene}. Basically, a more general interest in hexagonal structures arises from its non-separable character which implies the impossibility to generate a hexagonal lattice by a sum of two orthogonal one-dimensional lattices. 

\begin{figure}[t]
	\centering	
	\includegraphics[width=1.00\textwidth]{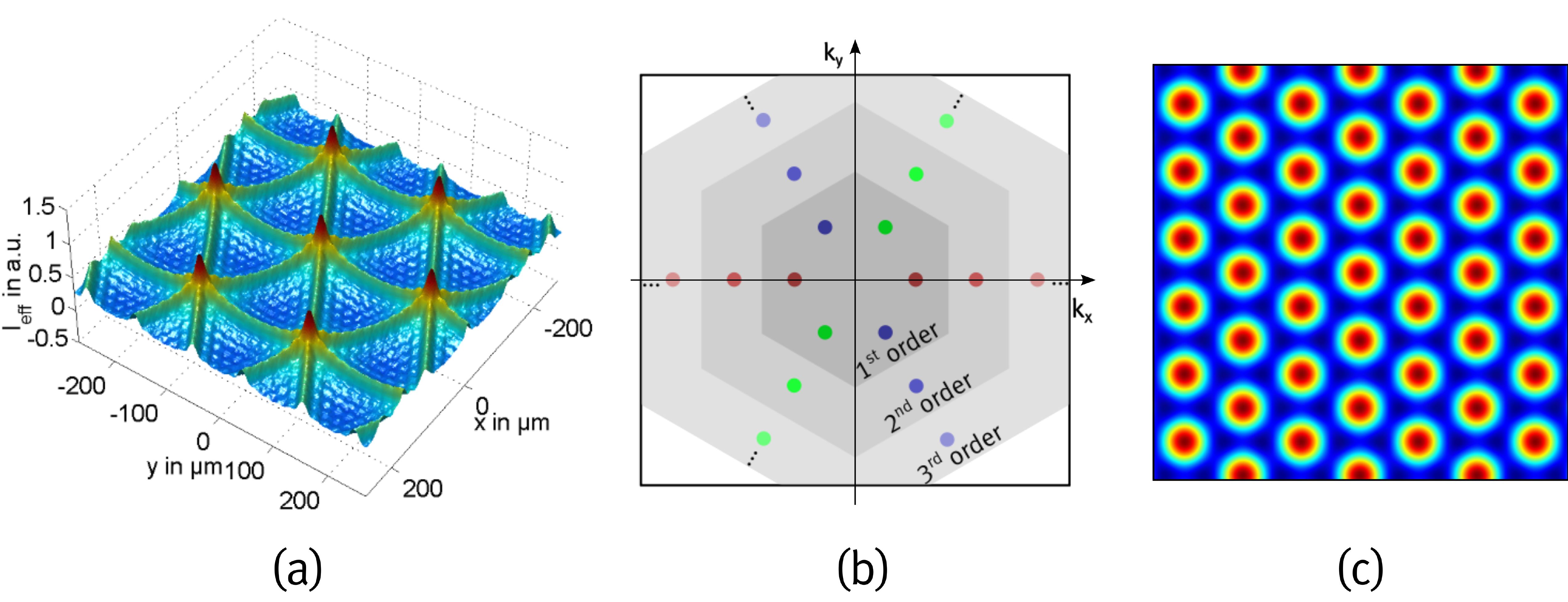}

	\caption{Schematics describing a multiperiodic 2D wire mesh lattice; (a): surface plot of the effective intensity calculated by series expansion up to an order of ten, (b): presentation of the relevant spatial frequencies, (c) periodic hexagonal lattice as the basis structure of each expansion term.}
	\label{fig:Fourier_wireMesh}
\end{figure}

In analogy to the considerations of Subsection \ref{ssec:2DRectStruct}, we start with the sum of 1D cosine lattices, resembling the hexagonal superstructure illustrated in \Fig{\ref{fig:Fourier_wireMesh}}(a):
\begin{eqnarray}\label{eq:hexagonal1D_basis}
\Hexa_{2D}(\vect r) = \sum_{m = 1}^{10} \frac{1}{m} \left[\frac{3}{2} + \cos\left(\vect k_{m}^1\vect r\right) + \cos\left(\vect k_m^2\vect r\right)+ \cos\left(\vect k_m^3\vect r\right)\right].
\end{eqnarray}
In this case, we need a hexagonal basis set of modulation vectors, which becomes apparent in the illustration of the relevant Fourier components in \Fig{\ref{fig:Fourier_wireMesh}}(b):
\begin{eqnarray}\label{eq:hexa_k_vectors}
\begin{split}
\vect k_{m}^j &= m\;\frac{2\pi}{g_1}\;\vect{e}_j = k_m\ \vect e_j \qquad \text{with}\  j = 1,2,3& \\ 
\text{and}&&\\
\vect{e}_1& = (\cos(\pi/6), \sin(\pi/6)),&\\  
\vect{e}_2& = (0, -1),&\\  
\vect{e}_3& = (-\cos(\pi/6), \sin(\pi/6)).&
\end{split}
\end{eqnarray}
Again we use the abbreviation $\phi_j$ with in this case $\phi_{j, m} = \vect{k}_m^j \vect r$ and $j = 1,2,3$ valid for all orders $m$. The term in square brackets appearing in \Eq{\ref{eq:hexagonal1D_basis}} and representing the unweighted intensity of every expansion order $m$ can be rewritten to
\begin{eqnarray}\label{eq:hex_squareField}
\left[\frac{3}{2} + \cos\left(\phi_1\right) + \cos\left(\phi_2\right)+ \cos\left(\phi_3\right)\right] = \frac{1}{2}\left[
3 + \e{i\phi_1} + \e{-i\phi_1} + \e{i\phi_2} + \e{-i\phi_2} + \e{i\phi_3} + \e{-i\phi_3}\right].
\end{eqnarray}
Now we have to find a 2D basis for the expansion, for instance given by the intensity of a hexagonal nondiffracting beam, which consists of three plane waves conforming the spatial frequency conditions of a nondiffracting beam (cf. Ref. \cite{discNondiffBeams}):
\begin{eqnarray} \label{eq:hex_squareField2}
\begin{split}
\left|\Psi_{3,0}(\vect r, k_m)\right|^2 =& \left| \e{i\varphi_1(\vect r, k_m)} + \e{i\varphi_2(\vect r, k_m)} + \e{i\varphi_3(\vect r, k_m)}\right|^2\\
=& 3+ \e{i(\varphi_1 - \varphi_2)} + \e{-i(\varphi_1 - \varphi_2)} + \e{i(\varphi_1 - \varphi_3)} + \e{-i(\varphi_1 - \varphi_3)} + \e{i(\varphi_2 - \varphi_3)} + \e{-i(\varphi_2 - \varphi_3)}
\end{split}
\end{eqnarray}
which can be identified with \Eq{\ref{eq:hex_squareField}} using $\phi_1 = \varphi_1 - \varphi_2, \phi_2 = \varphi_1 - \varphi_3, \phi_3 = \varphi_2 - \varphi_3$. Now, the arguments of the three plane waves in \Eq{\ref{eq:hex_squareField2}} are summations of the formerly introduced phase functions $\phi_j$.

Thus, the rule for the development of a multiperiodic 2D wire mesh function in hexagonal symmetry reads as
\begin{eqnarray}\label{eq:hexagonal2D}
\Hexa_{2D}(\vect r) = \sum_{m = 1}^{10}  \frac{1}{2m} \left|\Psi_{3,0}(\vect r, k_m)\right|^2
\end{eqnarray}
and is equivalent to \Eq{\ref{eq:hexagonal1D_basis}}.

In this connection, the fineness of the structure and thus the diameter of each spike is adaptable by broadening the spatial frequency bandwidth of the structure, which can be achieved by increasing the number of expansion orders. 

Besides the weighting of the fundamental structures, also the choice of the rotational symmetry brings out a parameter to vary the resulting multiperiodic structure. Thus the six-fold periodic structure presented in \Fig{\ref{fig:Fourier_wireMesh}}(c) is the basement for the hexagonal multiperiodic wire mesh structure presented in \Fig{\ref{fig:Fourier_wireMesh}}(a) which conserves the rotational symmetry.

\subsection{Ratchet superlattice}
In general, functional structures such as 2D ratchets have various intriguing applications in research fields such as biological microscopic machinery and Bose-Einstein
condensation \cite{quantumRatchet1}, as well as quantum mechanics in solid-state or atomic physics systems \cite{quantumRatchet2} due to their ability for long-distance mass transport caused by a low-amplitude potential. 

In analogy to the considerations regarding the two foregoing structures, we are going to develop the rule for the optical induction of a 2D ratchet structure. Therefore, we again start with a Fourier series expansion under the usage of a basis set of 1D fundamental lattices:
\begin{eqnarray}\label{eq:ratchet1D_basis}
\Ratch_{2D}(\vect r) = \sum_{m = 1}^{10} \frac{1}{m} \left[2 + \sin\left(\vect k_{m}^1\vect r\right) + \sin\left(\vect k_m^2\vect r\right)\right].
\end{eqnarray}
\begin{figure}[t]
	\centering	
	\includegraphics[width=1.00\textwidth]{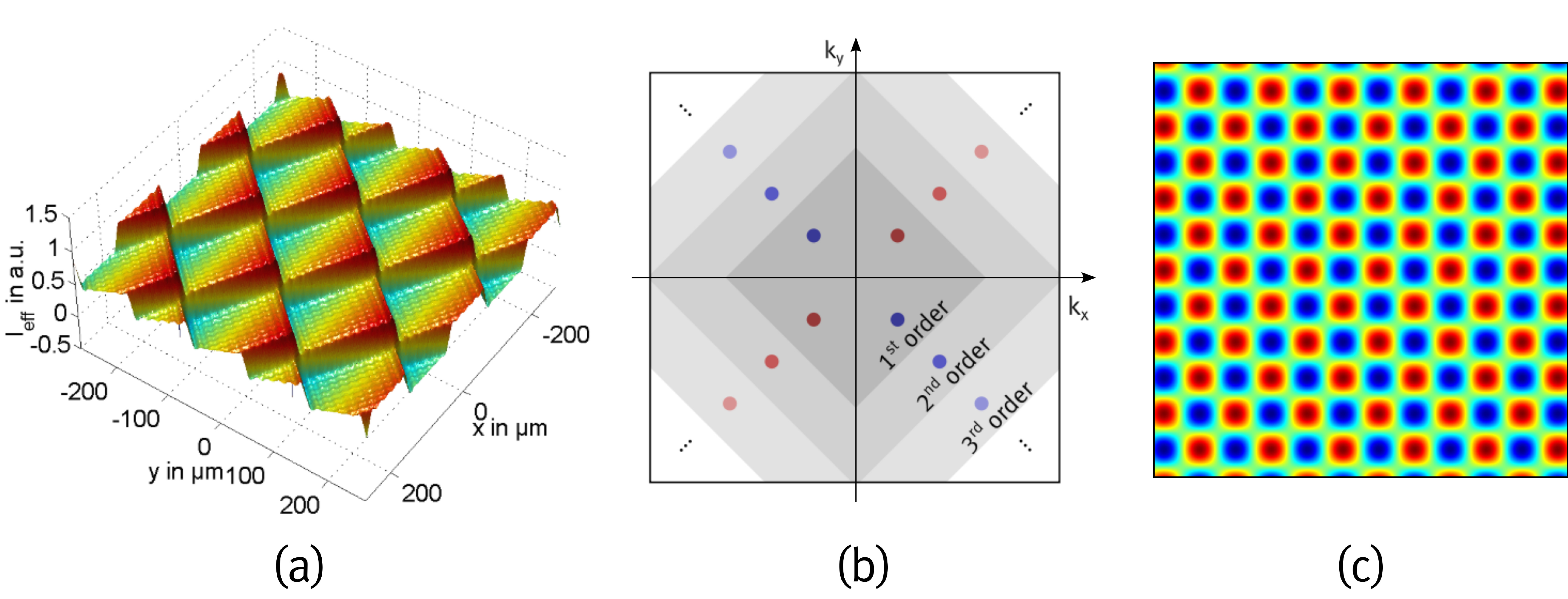}

	\caption{Schematics describing a multiperiodic 2D ratchet lattice; (a): surface plot of the effective intensity calculated by series expansion up to an order of ten, (b): presentation of the relevant spatial frequencies, (c) periodic square lattice as the basis structure of each expansion term.}
	\label{fig:Fourier_ratchet}
\end{figure}
Figure \ref{fig:Fourier_ratchet}(a) depicts the multiperiodic lattice in terms of the effective intensity distribution. Analog to \Fig{\ref{fig:Fourier_staircase}}(b) the corresponding series expansion respects specific spatial frequencies which are depicted in \ref{fig:Fourier_ratchet}(b) for the first three orders of expansion.

The set of modulation vectors assembles for each order $m$ from two mutual orthogonal vectors and are chosen to be
\begin{eqnarray}\label{eq:ratch_k_vectors}
\vect k_{m}^j = m\;\frac{2\pi}{g_1}\;\vect{e}_j = k_m\ \vect{e}_j\qquad \text{with}\  j = 1,2,\quad \vect{e}_1 = (1, 1),\  \vect{e}_2 = (1, -1).
\end{eqnarray}
With the previously introduced phase function $\phi_j = \vect k_j \vect r /2$, where $j = 1,2$, and under consideration of \Eq{\ref{eq:sinTerms}}, every expansion term of $\Ratch_{2D}$ can be expressed through the earlier introduced wave field $\Psi_{4,1}(\vect r, k_m)$, which again is depicted in \Fig{\ref{fig:Fourier_ratchet}}(c).
Hence, the series expansion for a multiperiodic 2D ratchet intensity distribution $\Ratch_{2D}$ composed of a 2D set of fundamental nondiffracting beam intensities is
\begin{eqnarray}\label{eq:ratchet2D}
\Ratch_{2D}(\vect r) = \sum_{m = 1}^{10}  \frac{1}{2m} \left|\Psi_{4,1}(\vect r, k_m)\right|^2,
\end{eqnarray}
where exclusively the weighting factor $1/(2m)$ as well as the relation between the modulation vectors of every order $m$ determines the difference between \Eq{\ref{eq:rectangular2D}} and \Eq{\ref{eq:ratchet2D}}.

\section{Experimental implementation} \label{sec:expImplmnt}
In the previous Section, we found the rules for the generation of multiperiodic structures deriving basic structures which resemble to intensities of nondiffracting beams. With these findings, we are able to experimentally implement corresponding photonic structures, where the induction procedure orientates on the theoretical series expansion mentioned in Eq. ({\ref{eq:rectangular2D}}), (\ref{eq:hexagonal2D}) and (\ref{eq:ratchet2D}). Here, the usage of 2D fundamentals bears an essential advantage regarding the temporal effort in comparison to 1D basic structures since for a series expansion of up to ten terms 30 1D lattice structures are needed, which differ among each other in modulation frequency (i.e. lattice period) and orientation, as well. Using 2D fundamentals, we only need 10 different multiplexing steps.

\begin{figure}[t]
	\centering	
	\includegraphics[width=.80\textwidth]{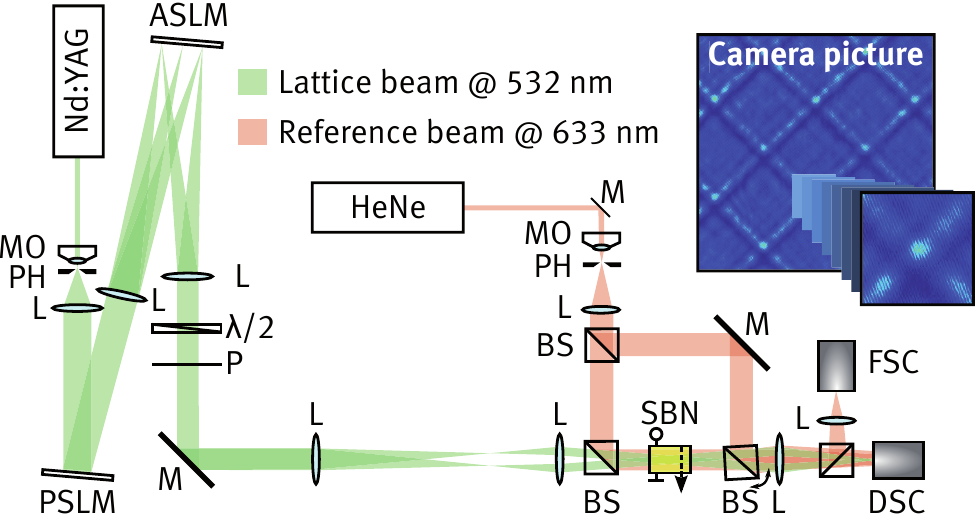}

	\caption{Schematic of experimental setup to optically induce and analyze spatial refractive index modulations of photorefractive crystals. Green colored light path represents the lattice inducing part, the probing part is colored in red. A/PSLM: amplitude/phase spatial light modulator, BS: beam splitter, D/FSC: direct/Fourier space camera, L: lens, $\lambda/2$: half-wave retardation plate, M: mirror, MO: microscope objective, P: polarizer, PH: pinhole, SBN: cerium-doped strontium barium niobate. Camera picture presents a typical recorded intensity pattern, magnification shows details of interference fringes.}
	\label{fig:Setup}
\end{figure}

The setup scheme for the experimental implementation of the optical induction of multiperiodic photonic structures is illustrated in \Fig{\ref{fig:Setup}}. An expanded continuous-wave solid-state laser beam with wavelength $\lambda = \unit{532}{\nano\meter}$ and an output power of approximately $\unit{80}{\milli\watt}$ is phase modulated by a computer controlled spatial light modulator (PSLM, {Holoeye} `Pluto', $1920\times 1080$ pixels). The phase pattern given to the PSLM and the displayed pattern of the following amplitude modulator for spatial frequency filtering (ASLM, {Holoeye} `LC-R~2500', $1024\times768$ pixels) are chosen to implement particular nondiffracting beams according to certain expansion terms. The phase distribution includes both amplitude as well as phase information about the desired wave field which is numerically calculated beforehand. Besides phase modulation, amplitude modulation can be achieved by an intensity weighted blazed grating to diffract the relevant light into a certain diffraction order \cite{PSLMAmpModulation}. In this context, the diffraction efficiency can be linearly controlled by adapting the slope of the blazed grating by means of a 2D weighting function in accordance with the calculated intensity distribution. Experimentally, the illumination time $t_m$ is easily adaptable via a short switching time of the PSLM.

By means of further optical elements, a desired wave field -- in our particular case a nondiffracting beam -- is imaged into a volume of interest. By implementing a demagnification factor of 6 we achieve an effective transverse illumination area of approximately $\unit{2.3 \times 1.3}{\milli \square \meter}$ at the imaging plane of the PSLM, the laser power over this area is about $\unit{300}{\micro\watt}$. Into this volume, whose longitudinal length extends several centimeters \cite{optInductionNdBs}, a photorefractive crystal is placed in order to modulate its refractive index. We employ a nonlinear cerium-doped strontium barium niobate (SBN) crystal that, externally biased, translates the intensity modulations into refractive index changes via the photorefractive effect. For the whole set of induction patterns, the external voltage is about \unit{800}{\volt} over a distance of $\unit{5}{\milli\meter}$ where the induction beam propagates along a $\unit{5}{\milli\meter}$ orientation of the crystal. The polarization of each writing wave field is chosen to be perpendicular to the symmetry axis of the crystal ($c$ axis) in order to minimize the influence of the induced refractive index change on the writing light due to a small electro-optical coefficient \cite{holoMuX}. Thus the refractive index of the medium can be assumed to be approximately homogeneous for this specific perpendicular state of polarization.

As a generalized approach, the temporal development of the refractive index change in a biased SBN crystal caused by an illumination pattern behaves exponentially \cite{incrMuX}, fades to saturation with the time constant $\tau_w$, and also gets exponentially erased ($\tau_e$) due to illumination with non-identical intensity distributions. This can be summarized by
\begin{align}\label{eq:RIDevelopment}
\begin{split}
\Delta n_w(t) &= \Delta n_{sat}\; [1 - \e{-t/\tau_w}],\\
\Delta n_e(t) &= \Delta n_0\; \e{-t/\tau_e},
\end{split}
\end{align}
with $\Delta n_0 = \Delta n(t = 0)$ and $\Delta n_{sat}$ as the saturation value of the refractive index change. In general, a written refractive index modulation is persistent in the SBN crystal until further illuminations.

In all multiperiodic expansions introduced in the previous section, the intensity belonging to a particular expansion order $m$ is weighted by a factor depending on $m$. This weighting factor is realized in the experiment by adapting the illumination time $t_m$ for each lattice-inducing beam.
Thereby, rather than controlling the external voltage or intensity of each fundamental structure, we regulate $t_m$ to weight a specific induction term. Hence, we determine the net illumination duration for one sequence of illumination to $t_\text{seq} = \sum_{m = 1}^{10} t_{m} = \unit{10}{\second}$ and execute about $20$ multiplexing sequences.

The optical path of the index-modulating part in \Fig{\ref{fig:Setup}} is colored in green, whereas the red-colored structure-analyzing part is subject of the next section. 

\section{Analysis of photonic structures} \label{sec:analysis}
In order to analyze the optically induced photonic structures, we implement a technique to retrieve the phase information of a probe wave field propagating through the induced photonic structure. Therefore, we apply digital-holographic techniques in terms of superimposing the probe beam with a reference plane wave of high spatial frequency, where a large angle enclosed by both wave vectors enables a high phase resolution. Using a common CMOS camera to record the phase-sensitive fringe pattern of the interference between the probe and the reference beam at the back face of the crystal allows to determine the whole probe wave field information rather than to merely detect the intensity distribution. In this way the phase retardation of the probe beam gives 2D resolved information about the refractive index structure and depth \cite{digiHolo}. In contrast to the writing process, the probe beam is polarized parallel to the $c$ axis to operate with a high electro-optical coefficient, which is approximately 6 times larger than the one for the writing procedure.
%\clearpage
\begin{figure}[thb]
	\centering	
	\includegraphics[width=.86\textwidth]{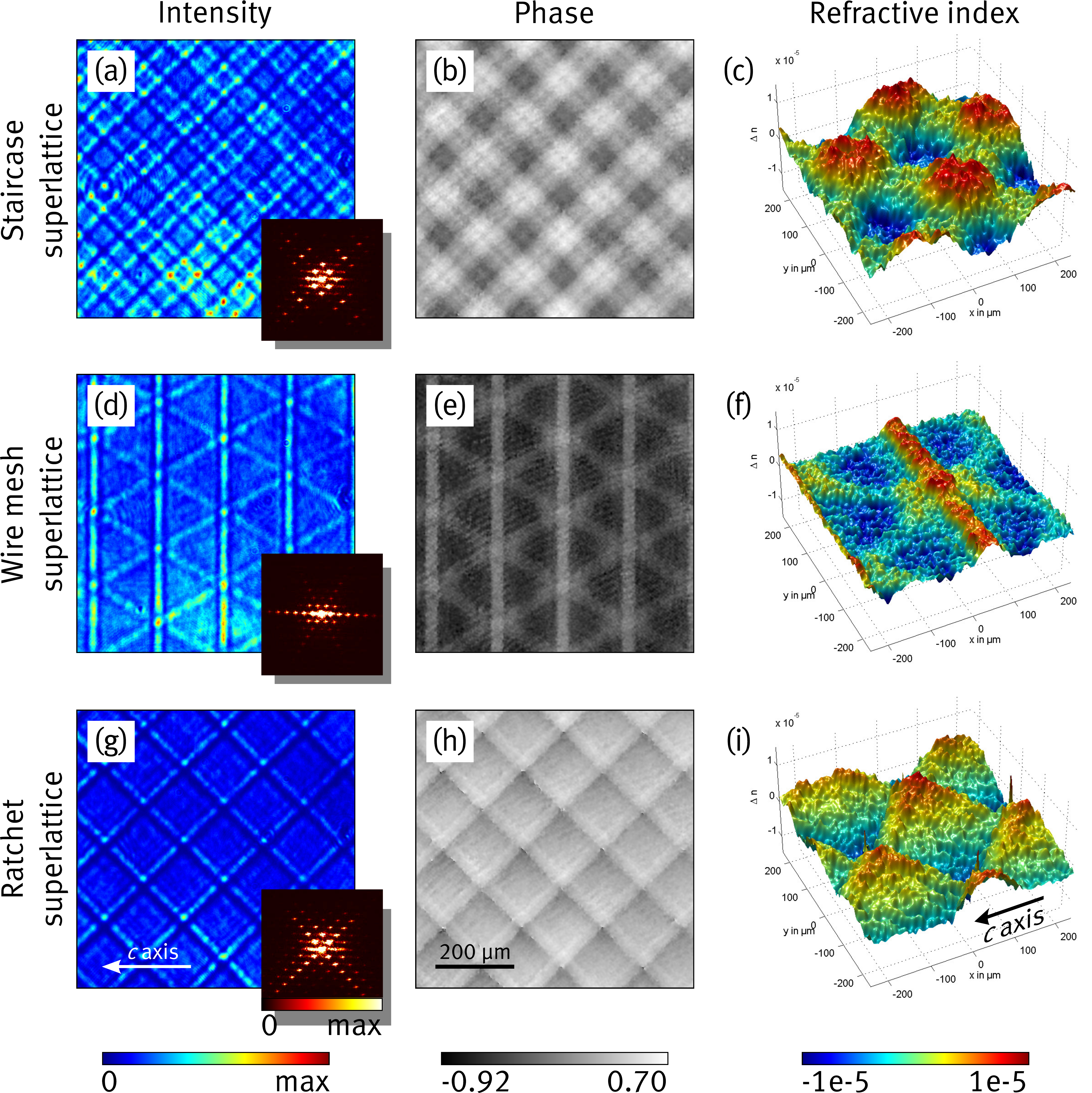}

	\caption{Analysis results of probing experiments using propagating plane wave of perpendicular incidence to crystal surface to characterize generated 2D photonic superlattices: (a)--(c) staircase structure, (d)--(f) wire mesh structure, (g)--(i) ratchet structure. Direction of $c$ axis is illustrated in (g) and (i), respectively. Scale bar for all real space pictures is appended in (h). Left column present intensity, middle column phase distribution of the plane probe wave in real space. Insets of left column pictures illustrate far field diffraction spectrum. Refractive index landscapes are depicted as surface plots in right column.}
	\label{fig:resultsRIStruct}
\end{figure}

To extract the phase information of the probe beam from the recorded interference pattern, one has to treat the data in analogy to an amplitude demultiplexing operation in electric signal processing. According to this, bandpass filtering in the vicinity of the carrier frequency followed by a frequency shift to zero provides the complex field information of the signal which is equal to the probe wave field and includes the desired phase information \cite{digiHolo2}. In order to bandpass filter the signal, we multiply the spatial frequency spectrum with a radial symmetric Hann window, whose center is the carrier frequency and its radius equals the half distance between carrier and zero frequency. In addition to a real space camera, we implemented a further camera to detect the intensity of the probe beam in the far field by usage of the Fourier-transforming attribute of an additional lens.

Figure \ref{fig:resultsRIStruct} presents the results of the analysis of the induced photonic staircase [(a)--(c)], hexagonal wire mesh [(d)--(f)] as well as ratchet structures [(g)--(i)]. Intensity as well as phase distributions of the probe plane wave in real space combined with the intensity distribution in the far field are illustrated in the lefthand and middle column, respectively. In order to determine the refractive index change of a particular area, we first identify the phase distribution of the homogeneous medium before the induction process was started. This distribution serves as the reference phase and is subtracted from the measured phase distribution of the induced structure. In this manner, exclusively the refractive index change $\Delta n$ caused by the optical induction process can be extracted by carrying out $\Delta n = \Delta \varphi \lambda / (2\pi d)$, where $d$ is the optical path length of the probe beam through the medium. 
The righthand column of \Fig{\ref{fig:resultsRIStruct}} visualizes in a surface plot the refractive index landscape of an area of approximately $\unit{0.16}{\milli\square\meter}$. 
%In addition to these illustrations, the induced structures are presented as animated surface plots supporting an enhanced spatial impression of all refractive index modulations (cf. Media 1, Media 2, Media 3).

Both the intensity and phase distribution of all three propagation measurements reveal the character of the respective structures introduced in Section \ref{sec:mpStruct}. The staircase superlattice presented in \Fig{\ref{fig:resultsRIStruct}}(a)--\Fig{\ref{fig:resultsRIStruct}}(c) is in great agreement with the simulations (cf. \Fig{\ref{fig:Fourier_staircase}}) showing three kinds of plateaus of distinct level where the intermediate level appears most [which becomes apparent in the surface plot of \Fig{\ref{fig:resultsRIStruct}}(c)]. The structure is thoroughly four-fold as the real-space results and the far field intensity distribution indicate, as well. According to this, the diffraction orders given in the inset of \Fig{\ref{fig:resultsRIStruct}}(a) resemble a discrete diffraction pattern multiplied by a 2D four-fold \textsl{sinc} function. We note that this specific rotational symmetry particularly results from the diamond orientation of the fundamental structures and is not the ordinary case for photorefractive media with drift dominated charge carrier transport processes causing an orientation anisotropy\cite{orientAniso}. 

\begin{figure}[t]
	\centering	
	\includegraphics[width=1\textwidth]{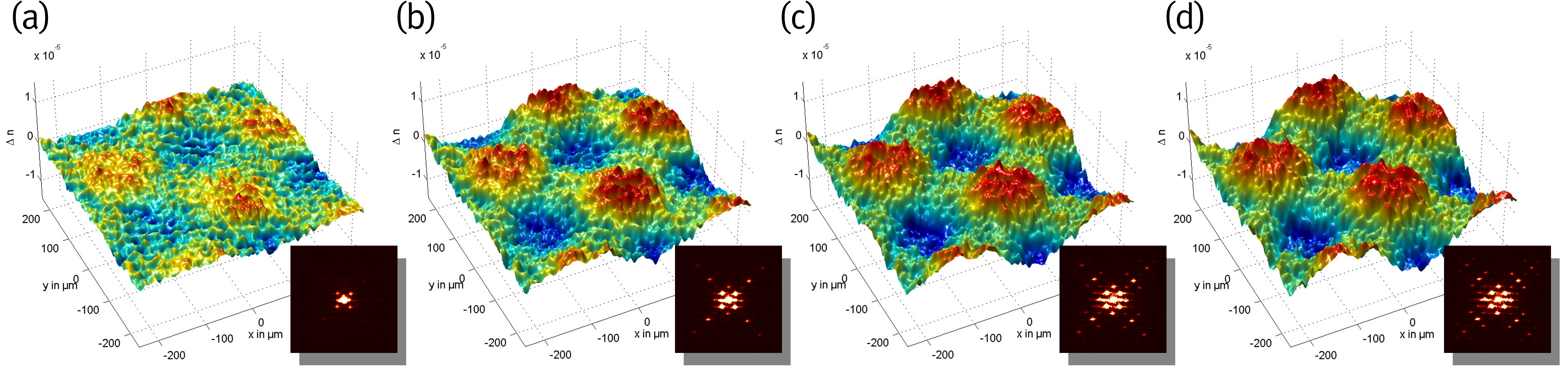}

	\caption{Time development of the rectangularly modulated refractive index structure. The detailed illumination times are (a)~$t_1 = \unit{20}{\second}$, (b)~$t_2 = \unit{80}{\second}$, (c)~$t_3 = \unit{140}{\second}$, (d)~$t_4 = \unit{200}{\second}$. Respective insets present corresponding intensity distributions in far field.}
	\label{fig:RI_development}
\end{figure}

In contrast to the staircase superlattice, such an anisotropy becomes apparent for the hexagonal wire mesh structure where intensity and phase distribution [cf. \Fig{\ref{fig:resultsRIStruct}}(d)--\Fig{\ref{fig:resultsRIStruct}}(f)] display a stronger refractive index modulation along the $c$ axis compared to any other direction. Thus, vertical lines perpendicular to the $c$ axis are more pronounced than the ones that are inclined by \unit{30}{\degree} to the $c$ axis. The far field picture presented in the inset of \Fig{\ref{fig:resultsRIStruct}}(d) and the surface plot in \Fig{\ref{fig:resultsRIStruct}}(f) emphasize the influence of the crystal's orientation anisotropy, additionally. 

With the ratchet superlattice presented in \Fig{\ref{fig:resultsRIStruct}}(g)--\Fig{\ref{fig:resultsRIStruct}}(i), we introduce a further an\-iso\-tropy-unaffected structure besides the staircase superlattice, though this structure is not rotational symmetric as the straightering character is inherent for the ratchet structure. In particular, the far field intensity illustrated in the inset of \Fig{\ref{fig:resultsRIStruct}}(g) manifests a preferred direction of light diffraction (spectral quarter above center). Moreover, in contrast to the 1D ratchet structure presented in \cite{ratch1D}, the 2D ratchet superlattice can be oriented in four different ways without disturbance by the anisotropy and can further be used to direct light parallelly as well as antiparallelly or perpendicularly up as well as down to the $c$ axis, respectively. Altogether, especially the ratchet structure is preferred to be applied as a light director and marks a throughout relevant system for propagation experiments in the linear and in the nonlinear regime, as well. 

In the setup presented in \Fig{\ref{fig:Setup}} we use two distinctive wavelengths -- $\unit{532}{\nano\meter}$ for the induction of the refractive index modulation (frequency doubled solid state Nd:YAG cw laser) and $\unit{633}{\nano\meter}$ as the probe beam wavelength (He-Ne cw laser). The advantage of using red laser light in our probing experiments is the smaller sensitivity of SBN in this wavelength regime compared to wavelengths in the green regime around $\unit{532}{\nano\meter}$ \cite{holoMuX}.

By using a separate probe laser, we are able to timely-resolved detect the refractive index modulation, which facilitates the development of the index change due to optical induction. Figures \ref{fig:RI_development}(a)--\ref{fig:RI_development}(d) present the developing index structure of a fixed area for four different points in time $t_1 = \unit{20}{\second}$, $t_2 = \unit{80}{\second}$, $t_3 = \unit{140}{\second}$, and $t_4 = \unit{200}{\second}$. The formation of the staircase photonic structure becomes obvious already after $t_3 = \unit{140}{\second}$, and also the far field intensity reveals characteristic details of the desired structure after several seconds of illumination.
							
\section{Conclusion}
In conclusion, we introduced a ubiquitous method to generate 2D photonic superstructures by use of 2D nondiffracting beams. These writing beams serve as fundamentals of an optical series expansion to generate multiperiodic photonic structures. In such a manner, e.g. anisotropic, graphene-like, as well as discontinuous functional structures are realizable. Although we proved a successful induction of three representative complex superlattices in a photorefractive SBN crystal, the technique is easily transferable to all photosensitive media, enabling the development of photonic structures of long expansion in transverse as well as in longitudinal direction. Also the implementation of multiperiodic optical atom or particle traps is one prospective application of this technique. In general, our method offers a highly-functional tool for the generation of two-dimensional multiperiodic systems, providing the basement of the investigation of so far unexplored fascinating linear as well as nonlinear light propagation effects.


\begin{thebibliography}{99}
\small
\bibitem{BandStructures} E.~Yablonovitch, T.~J.~Gmitter, and K.~M.~Leung, ``Photonic band structure: The face-centered-cubic case employing nonspherical atoms,'' Phys. Rev. Lett. {\bf 67,} 2295-2298 (1991).

\bibitem{BraggReflection} M.~G.~Moharam, T.~K.~Gaylord, and R.~Magnusson, ``Bragg diffraction of finite beams by thick gratings,'' J. Opt. Soc. Am. {\bf 70,} 300-304 (1980).

\bibitem{BlochOscillation} R.~Sapienza, P.~Costantino, D.~Wiersma, M.~Ghulinyan, C.~J.~Oton, and L.~Pavesi, ``Optical analogue of electronic Bloch oscillations,'' Phys. Rev. Lett. {\bf 91,} 263902 (2003).

\bibitem{BlochZener} H.~Trompeter, W.~Krolikowski, D.~N.~Neshev, A.~S.~Desyatnikov, A.~A.~Sukhorukov, Y.~S.~Kivshar, T.~Pertsch, U.~Peschel, and F.~Lederer, ``Bloch oscillations and Zener tunneling in two-dimensional photonic lattices,'' Phys. Rev. Lett. {\bf 96,} 053903 (2006).

\bibitem{AndersonLocalPeriod} T.~Schwartz, G.~Bartal, S.~Fishman, and M.~Segev, ``Transport and Anderson localization in disordered two-dimensional photonic lattices,'' Nature {\bf 446,} 52-55 (2007).

\bibitem{discSoliton1} D.~N.~Christodoulides and R.~I.~Joseph, ``Discrete self-focusing in nonlinear arrays of coupled waveguides,'' Opt. Lett. {\bf 13,} 794-796 (1988).

\bibitem{discSoliton2a} D.~Neshev, E.~Ostrovskaya, Y.~Kivshar, and W.~Krolikowski, ``Spatial solitons in optically induced gratings,'' Opt. Lett. {\bf 28,} 710-712 (2003).

\bibitem{discSoliton2b} J.~W.~Fleischer, M.~Segev, N.~K.~Efremidis, and D.~N.~Christodoulides, ``Observation of two-dimensional discrete solitons in optically induced nonlinear photonic lattices,'' Nature {\bf 422,} 147-150 (2003).

\bibitem{vortSoliton1} O.~Manela, O.~Cohen, G.~Bartal, J.~W.~Fleischer, and M.~Segev, ``Two-dimensional higher-band vortex lattice solitons,'' Opt. Lett. {\bf 29,} 2049-2051 (2004).
\bibitem{vortSoliton2} G.~Bartal, O.~Manela, O.~Cohen, J.~W.~Fleischer, and M.~Segev, ``Observation of second-band vortex solitons in 2D photonic lattices,'' Phys. Rev. Lett. {\bf 95,} 053904 (2005).

\bibitem{multVortSoliton1} T.~J.~Alexander, A.~S.~Desyatnikov, and Y.~S.~Kivshar, ``Multivortex solitons in triangular photonic lattices,'' Opt. Lett. {\bf 32,} 1293-1295 (2007).
\bibitem{multVortSoliton2} B.~Terhalle, T.~Richter, A.~S.~Desyatnikov, D.~N.~Neshev, W.~Krolikowski, F.~Kaiser, C.~Denz, and Y.~S.~Kivshar, ``Observation of multivortex solitons in photonic lattices,'' Phys. Rev. Lett. {\bf 101,} 013903 (2008).

\bibitem{quasiperiodPhotonStructures1} Y.~S.~Chan, C.~T.~Chan, and Z.~Y.~Liu, ``Photonic band gaps in two dimensional photonic quasicrystals,'' Phys. Rev. Lett. {\bf 80,} 956-959 (1998).

\bibitem{quasiperiodPhotonStructures2} B.~Freedman, G.~Bartal, M.~Segev, R.~Lifshitz, D.~N.~Christodoulides, and J.~W.~Fleischer, ``Wave and defect dynamics in nonlinear photonic quasicrystals,'' Nature {\bf 440,} 1166-1169 (2006).

\bibitem{BesselSolitons1a} Y.~V.~Kartashov, A.~A~Egorov, V.~A.~Vysloukh, and L.~Torner, ``Stable soliton complexes and azimuthal switching in modulated Bessel optical lattices,'' Phys. Rev. E {\bf 70,} 065602(R) (2004).

\bibitem{BesselSolitons1b}Y.~V.~Kartashov, V.~A.~Vysloukh, and L.~Torner, ``Rotary solitons in bessel optical lattices,'' Phys. Rev. Lett. {\bf 93,} 093904 (2004).

\bibitem{BesselSolitons1c}Y.~V.~Kartashov, V.~A.~Vysloukh, and L.~Torner, ``Soliton shape and mobility control in optical lattices,'' Prog. Optics {\bf 52,} 63-148 (2009).

\bibitem{BesselSolitons2} S.~Huang, P.~Zhang, X.~Wang, and Z.~Chen, ``Observation of soliton interaction and planetlike orbiting in Bessel-like photonic lattices,'' Opt. Lett. {\bf 35,} 2284-2286 (2010).

\bibitem{MathieuSolitons} F.~Ye, D.~Mihalache, and B.~Hu, ``Elliptic vortices in composite Mathieu lattices,'' Phys. Rev. A {\bf 79,} 053852 (2009).

\bibitem{AndersonLocalization} D.~M.~Jovi\'c, M.~R.~Beli\'c, and C.~Denz, ``Anderson localization of light at the interface between linear and nonlinear dielectric media with an optically induced photonic lattice,'' Phys. Rev. A {\bf 85,} 031801(R) (2012).

\bibitem{KleinTunneling} S.~Longhi, ``Klein tunneling in binary photonic superlattices,'' Phys. Rev. B {\bf 81,} 075012 (2010).

\bibitem{Zitterbewegung1} S.~Longhi, ``Photonic analog of Zitterbewegung in binary waveguide arrays,'' Opt. Lett. {\bf 35,} 235-237 (2010).
\bibitem{Zitterbewegung2} F.~Dreisow, M.~Heinrich, R.~Keil, A.~T\"unnermann, S.~Nolte, S.~Longhi, and A.~Szameit, ``Classical simulation of relativistic Zitterbewegung in photonic lattices,'' Phys. Rev. Lett. {\bf 105,} 143902 (2010).

\bibitem{superlattSoliton} M.~Heinrich, Y.~V.~Kartashov, L.~P.~R.~Ramirez, A.~Szameit, F.~Dreisow, R.~Keil, S.~Nolte, A.~T\"unnermann, V.~A.~Vysloukh, and L.~Torner, ``Observation of two-dimensional superlattice solitons,'' Opt. Lett. {\bf 34,} 3701-3703 (2009).

\bibitem{directLWrite} M.~Heinrich, R.~Keil, F.~Dreisow, A.~T\"unnermann, A.~Szameit, and S.~Nolte, ``Nonlinear discrete optics in femtosecond laser-written photonic lattices,'' Appl. Phys. B {\bf 104,} 469-480 (2011).

\bibitem{lithography} M.~Campbell, D.~N.~Sharp, M.~T.~Harrison, R.~G.~Denning, and A.~J.~Turberfield, ``Fabrication of photonic crystals for the visible spectrum by holographic lithography,'' Nature {\bf  404,} 53-56 (2000).

\bibitem{prMedium1} B.~Terhalle, A.~S.~Desyatnikov, D.~N.~Neshev, W.~Krolikowski, C.~Denz, and Y.~S.~Kivshar, ``Dynamic diffraction and interband transitions in two-dimensional photonic lattices,'' Phys. Rev. Lett. {\bf  106,} 083902 (2011).

\bibitem{prMedium2} A.~S.~Desyatnikov, N.~Sagemerten, R.~Fischer, B.~Terhalle, D.~Tr\"ager, D.~N.~Neshev, A.~Dreischuh, C.~Denz, W.~Krolikowski, and Y.~S.~Kivshar, ``Two-dimensional self-trapped nonlinear photonic lattices,'' Opt. Express {\bf  14,} 2851-2863 (2006).

\bibitem{optInductionNdBs} P.~Rose, M.~Boguslawski, and C.~Denz, ``Nonlinear lattice structures based on families of complex nondiffracting beams,'' New J. Phys. {\bf 14,} 033018 (2012).

\bibitem{optInd3DLatt} J.~Xavier, M.~Boguslawski, P.~Rose, J.~Joseph, and C.~Denz, ``Reconfigurable optically induced quasicrystallographic three-dimensional complex nonlinear photonic lattice structures,'' Adv. Mater. {\bf 22,} 356-360 (2010).

\bibitem{ndBs1} J.~Durnin, ``Exact solutions for nondiffracting beams. I. The scalar theory,'' J. Opt. Soc. Am. A {\bf 4,} 651-654 (1987).

\bibitem{ndBs2} Z.~Bouchal, ``Nondiffracting optical beams: Physical properties, experiments, and applications,'' Czech. J. Phys. {\bf 53,} 537-578 (2003).

\bibitem{ndBs3} M.~A.~Bandres, J.~C.~Guti\'errez-Vega, and S.~Ch\'avez-Cerda, ``Parabolic nondiffracting optical wave fields,'' Opt. Lett. {\bf 29,} 44-46 (2004).

\bibitem{MuXDataStorage} C.~Denz, G.~Pauliat, G.~Roosen, and T.~Tschudi, ``Volume hologram multiplexing using a deterministic phase encoding method,'' Opt. Commun. {\bf 85,} 171-176 (1991).

\bibitem{incrMuX} Y.~Taketomi, J.~E.~Ford, H.~Sasaki, J.~Ma, Y.~Fainman, and S.~H.~Lee ``Incremental recording for photorefractive hologram multiplexing,'' Opt. Lett. {\bf 16,} 1774-1776 (1991).

\bibitem{holoMuX} P.~Rose, B.~Terhalle, J.~Imbrock, and C.~Denz, ``Optically induced photonic superlattices by holographic multiplexing,'' J. Phys. D: Appl. Phys. {\bf 41,} 224004 (2008).

\bibitem{ratch1D} M.~Boguslawski, A.~Kelberer, P.~Rose, and C.~Denz, ``Photonic ratchet superlattices by optical multiplexing,'' Opt. Lett. {\bf 37,} 797-799 (2012).

\bibitem{vortexBeams} J.~Becker, P.~Rose, M.~Boguslawski, and C.~Denz, ``Systematic approach to complex periodic vortex and helix lattices,'' Opt. Express {\bf 19,} 9848-9862 (2011).

\bibitem{discNondiffBeams} M.~Boguslawski, P.~Rose, and C.~Denz, ``Increasing the structural variety of discrete nondiffracting wave fields,'' Phys. Rev. A {\bf 84,} 013832 (2011).

\bibitem{graphene} A.~K.~Geim and K.~S.~Novoselov, ``The rise of graphene,'' Nat. Mater. {\bf 6,} 183-191 (2007).

\bibitem{quantumRatchet1} T.~Salger, S.~Kling, T.~Hecking, C.~Geckeler, L.~Morales-Molina, and M.~Weitz, ``Directed transport of atoms in a Hamiltonian quantum ratchet,'' Science {\bf 326,} 1241-1243 (2009).

\bibitem{quantumRatchet2} P. H\"anggi and F. Marchesoni, ``Artificial Brownian motors: Controlling transport on the nanoscale,'' Rev. Mod. Phys. {\bf 81,} 387-442 (2009).

\bibitem{PSLMAmpModulation} J.~A.~Davis, D.~M.~Cottrell, J.~Campos, M.~J.~Yzuel, and I.~Moreno ``Encoding amplitude information onto phase-only filters,'' Appl. Opt. {\bf 38,} 5004-5013 (1999).

\bibitem{digiHolo} B.~Kemper and G.~von~Bally, ``Digital holographic microscopy for live cell applications and technical inspection,'' Appl. Opt. {\bf 47,} A52-A61 (2008).

\bibitem{digiHolo2} U.~Schnars and W.~Jueptner, \textit{Digital Holography} (Springer, 2005).

\bibitem{orientAniso} B.~Terhalle, A.~S.~Desyatnikov, C.~Bersch, D.~Tr\"ager, L.~Tang, J.~Imbrock, Y.~S.~Kivshar, and C.~Denz, ``Anisotropic photonic lattices and discrete solitons in photorefractive media,'' Appl. Phys. B {\bf 86,} 399-405 (2007).

\end{thebibliography}
\end{document}